\documentclass [aps,prb,twocolumn,amssymb,amsmath,showpacs,floatfix,superscriptaddress]{revtex4-2}

 \def\Xint#1{\mathchoice
   {\XXint\displaystyle\textstyle{#1}}%
   {\XXint\textstyle\scriptstyle{#1}}%
   {\XXint\scriptstyle\scriptscriptstyle{#1}}%
   {\XXint\scriptscriptstyle\scriptscriptstyle{#1}}%
   \!\int}
\def\XXint#1#2#3{{\setbox0=\hbox{$#1{#2#3}{\int}$}
     \vcenter{\hbox{$#2#3$}}\kern-.5\wd0}}

\def\dashint{\Xint-}

\usepackage{amsmath}
\usepackage{amssymb}
\usepackage{amsfonts}
\usepackage{graphicx}
\usepackage{verbatim}
\usepackage{bm}
\usepackage{mathbbol}
\usepackage[dvipsnames]{xcolor}
\setcitestyle{compress}

\usepackage{mathtools}
\usepackage{esint}

\usepackage{tabularx}
 
\usepackage[normalem]{ulem}

\usepackage{graphicx,epsfig,amsfonts,amssymb}
\usepackage{bm}
\usepackage{times}
\usepackage{lipsum}
\usepackage{verbatim}

\usepackage{hyperref}
\hypersetup{
    colorlinks,
    citecolor=blue,
    filecolor=blue,
	    linkcolor=blue,
    urlcolor=blue
}

\newcommand{\beg}{\begin{equation}}
\newcommand{\en}{\end{equation}}

 \newcommand{\lam}{\lambda}

\newcommand{\eref}[1]{Eq.~(\ref{#1})}
\newcommand{\re}[1]{(\ref{#1})}

\newcommand{\esref}[1]{Eqs.~(\ref{#1})}
\renewcommand{\Re}{\mathrm{Re}}
\renewcommand{\Im}{\mathrm{Im}}
\newcommand{\Tr}{\mathrm{Tr}}

\begin{document}

\title{Instability of Metals  with Respect to Strong Electron-Phonon Interaction}

\author{Emil A. Yuzbashyan}
\affiliation{Department of Physics and Astronomy, Center for Materials Theory, Rutgers University, Piscataway, New Jersey 08854, USA}

\author{Boris L. Altshuler}
\affiliation{Physics Department, Columbia University, 538 West 120th Street, New York, New York 10027, USA}

\author{Aniket Patra}
\affiliation{Center for Theoretical Physics of Complex Systems, Institute for Basic Science, Daejeon 34126, Republic of Korea}

\begin{abstract}

 We show that thermal equilibrium between conduction electrons and phonons becomes kinetically unstable when the renormalized electron-phonon coupling exceeds a certain threshold. We prove that negative electronic specific heat, $C_\mathrm{el} < 0$, is sufficient to trigger the instability. Specifically, the instability sets in as soon as the quasiparticle weight becomes negative over a range of energies, even before $C_\mathrm{el}$ turns negative. This is an inherently nonequilibrium phenomenon, occurring prior to the formation of any equilibrium phase. Depending on the system, it can proceed along different pathways, ultimately resulting in a structural transition to an insulating or metallic state.
 
\end{abstract}

\maketitle

 Although the idea that  the strength of the \textsl{physical} (renormalized) electron-phonon  interaction in metals is somehow limited by  a lattice instability has been around for more than half a century, a clear understanding of whether a    fundamental  upper bound of this kind truly exists  is sill lacking~\cite{anderson,cohen,kresin}.
 On one hand, it is well known that the  metallic phase of the electron-phonon models, such as Holstein, Fr\"olich, Su-Schrieffer-Heeger, etc., becomes unstable when the \textsl{bare} (unrenormalized) dimensionless electron-phonon coupling $\lam_0>0.5$~\cite{migdal,eli1st,agd,ZSRBMPS}. The instability is variably described in the literature as  lattice reconstruction or charge-density-wave     or as polaronic or bipolaronic transition~\cite{ZSRBMPS,Chakraverty,scalettar,kabanov,roland,ono,alexandrov,capone,esterlis,scalapino,Nosarzewski,Bradley,polaron}. On the other hand, the physical   electron-phonon coupling $\lam$, which determines the superconducting $T_c$ and other physical properties, diverges at the instability~\cite{migdal,eli1st,agd}.  Since  $T_c \propto\sqrt{\lam}$ for large $\lam$~\cite{allendynes,mitrovic,carbotte,combescot,michael1,michael2},  $T_c$ too can   take arbitrarily large values.

 The situation is further complicated by the fact that the electron-phonon models cannot be consistently derived from the underlying  Hamiltonian of electrons and ions interacting via Coulomb forces and, therefore, do not provide a fully accurate  description~\cite{kagan,geilikman,karakozov,Sadovskii,Marini,berges}.
When supplemented with the actual phonon spectrum, they  do    capture the effects due to  phonon mediated electron-electron interaction~\cite{kagan,geilikman,lee1_main,karakozov,Sadovskii}. Yet it is impossible to obtain the correct phonon spectrum from within these models. The main reason is  double counting (also known as overscreening)   electronic polarization in the renormalized phonon spectrum~\cite{kresin,kagan,geilikman,Marini,berges}. This contribution is already taken into account once when obtaining, e.g., the Holstein or Fr\"olich models from the   electron-ion Hamiltonian, and the subsequent calculation of the phonon spectrum  within these models includes  it once more.  As a result, physical phonon frequencies $\omega_{\bm k}\approx \omega_{\bm k 0}\sqrt{1-2\lam_0}$, where $\omega_{\bm k 0}$ are the bare frequencies, soften at $\lam_0=0.5$. A \textsl{spurious} lattice instability, absent in  the  original electron-ion Hamiltonian, develops, while the renormalized electron-phonon interaction constant $\lam=\frac{\lam_0}{1-2\lam_0}$ diverges as $\lam_0\to 0.5$.  In many cases, the polaronic and charge-density-wave (P\slash CDW)  instabilities  found in the electron-phonon models is simply this artificial phonon softening~\cite{polaron}.   
  
 Most importantly, regardless of whether or not we accept the P\slash CDW instability at $\lam_0\approx 0.5$ as real, we see that it does not lead to  well-defined  upper bounds on $\lam$
and   $T_c$. 

In this Letter, we identify a \textsl{kinetic instability in metals} that sets a strict upper bound on the \textsl{physical} electron-phonon interaction strength. This instability is \textsl{fundamentally distinct} from the P\slash CDW instability also because it is a \textsl{nonequilibrium phenomenon}, does \textsl{not} involve phonon softening, and is \textsl{not} based on effective electron-phonon models.
Previous work demonstrated that the electronic specific heat, $C_\mathrm{el},$ becomes negative when $\lam$ exceeds a certain threshold~\cite{breakdown,zhang2,semenok}. However, since electrons are strongly coupled to phonons, it is not immediately obvious that this necessarily leads to an instability. Additionally, because the phonon contribution to the total specific heat is much larger than $|C_\mathrm{el}|$~\cite{regime}, the overall specific heat remains positive~\cite{breakdown,zhang2}.

Here, we \textsl{prove rigorously} that thermal equilibrium between electrons and phonons becomes \textsl{intrinsically unstable whenever $C_\mathrm{el}<0$}. More importantly, we find that metals \textsl{undergo a kinetic instability even before $C_\mathrm{el}$} turns 
negative---this occurs as soon as the quasiparticle weight, $Z_0(E)$ (also known as the wave function renormalization constant), becomes  negative over a finite energy range. This is an \textsl{absolute instability}, meaning the system is unstable to \textsl{infinitesimal perturbations} around equilibrium, and it is a fundamental feature of any fermionic system with a retarded interaction mediated by virtual bosons.
Instabilities of this kind  signal a \textsl{first-order phase transition}---for example, the negative compressibility of a van der Waals gas indicates a liquid-gas transition~\cite{epstein,callen}. Here, we anticipate that the system undergoes a \textsl{structural transition} into an insulator or another metal~\cite{breakdown,semenok}.

\begin{figure*}[h!tb]
\includegraphics[width=.96\textwidth]{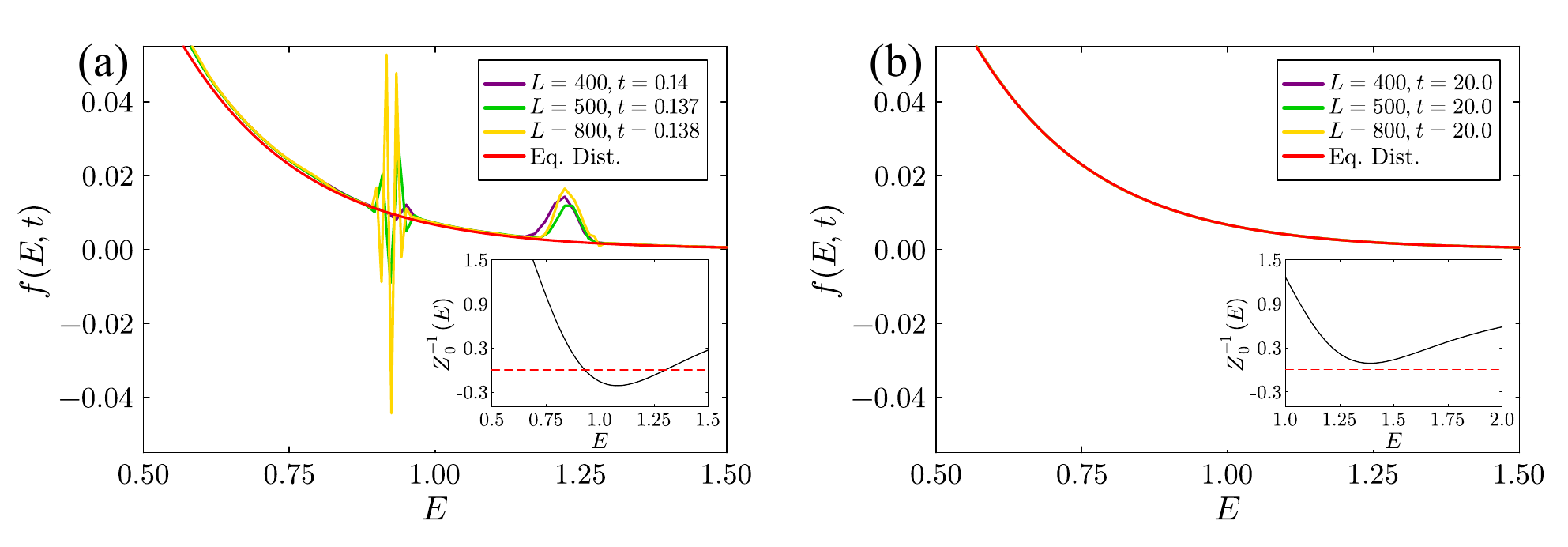}
\caption{ Instability in the full kinetic equation. We simulate  \eref{ke} for Einstein phonons with  (a) $\lambda  = 2.169>\lam_c=1.27$ and (b) $\lam=1.0<\lam_c$.  The initial electron temperature  $T_\textrm{in} = 0.202$ is slightly above the phonon temperature $T_{0} = 0.20$.     Insets show the corresponding inverse quasiparticle weight $Z^{-1}_0 \left( E \right)$. The solid red line in both panels is the  equilibrium Fermi distribution $f_0(T_0)$. The unstable growth of the electron distribution function $f(E,t)$ in  (a) is most prominent near the energies $E = 0.94$ and $1.30$ satisfying $Z^{-1}_0 (E) = 0.$ Because of the highly oscillating behavior,   the simulation can run only for a short time in (a).   There  is no instability in (b), since  $Z^{-1}_0 (E) > 0$, and the solution converges to  $f_0(T_0)$.}
\label{FigInstability}
\end{figure*}

The kinetic equation for the electron distribution function $f(E, t)$ in the normal state in the presence of  electron-phonon interactions is~\cite{prange1,rammer}
\beg
 (1-\Sigma') \dot f +f' \dot\Sigma = I_\mathrm{ep},
\label{ke}
\en
where the dot and prime denote derivatives with respect to time  and energy. $I_\mathrm{ep}$  is the electron-phonon    collision integral,
\beg
\begin{split}
I_\mathrm{ep}=2\pi\! \int_0^\infty \!\!\! d\omega \alpha^2 F(\omega)\bigr[ \bigl.& N_0(T_0)(f_++f_--2f) \\
& \bigl.- f(f_+-f_-) +f_+-f\bigr].
\end{split}
\en
Here $f_\pm=f(E\pm\omega)$, and
\beg
 \alpha^2 F(\omega) =\sum_k \frac{\lam_k\omega_k}{2} \delta(\omega-\omega_k)
\en
is the Eliashberg function.   All phonon frequencies satisfy $\omega_k > 0$, so our treatment does not involve phonon softening. Since the electron-phonon interaction enters through a Fermi surface average in $\alpha^2 F(\omega)$, its detailed momentum dependence plays no role in the electron dynamics at leading order in $\Omega / E_F$, where $E_F$ is the Fermi energy and $\Omega$ the maximum phonon frequency.

The electron self-energy is
\beg
\Sigma[f]= \dashint d x \int_0^\infty \!\!\! d\omega \alpha^2 F(\omega) \frac{ f(x+\omega) -f(x-\omega)}{E-x},
\label{se1}
\en
  $N_0(T_0)=[e^{\frac{\omega}{T_0}}-1]^{-1}$ is the Bose distribution at temperature $T_0$,
 $\omega_k$ are   phonon frequencies, and $\lam_k$ are the coupling constants.
This kinetic equation has been derived within the same approximation as the Migdal-Eliashberg (ME) theory~\cite{seesup}. 
Its r.h.s. explicitly takes into account the electron self-energy due to the electron-phonon interaction, while the   direct electron-electron interactions are included through the effective parameters of the underlying Fermi liquid. 
We also took spatially uniform initial conditions, so that the electron and phonon distribution functions are independent of the spatial coordinates, and assumed that phonons are in thermal   equilibrium.  The latter is an excellent approximation,  because $\frac{C_\mathrm{ph} }{|C_\mathrm{el}|}\sim \frac{E_F}{\Omega}$~\cite{regime}.  Since the small parameter in the ME theory is $\frac{\Omega}{E_F}\to0$, $C_\mathrm{ph}\to\infty$ in the limit where this theory is exact. This means that phonons act as a thermal bath with an infinite heat capacity for the fermions~\cite{seesup}. 

The fixed point of \eref{ke} is the equilibrium Fermi distribution $f_0(E, T_0)=[e^{\frac{E}{T_0}}+1]^{-1}$. In equilibrium, $1-\Sigma'[f_0]\equiv Z_0^{-1}(E)$ has the meaning of the inverse quasiparticle weight function. Explicitly~\cite{app},
\beg
Z_0^{-1}=1-(2\pi T_0)^{-1}  \int \!\! d\omega \alpha^2 F(\omega) \Im(\psi^{(1)}_+ -\psi^{(1)}_-),
\label{zfull}
\en
where $\psi^{(1)}_\pm=\psi^{(1)}\left(\frac{1}{2}+\frac{iE\pm i\omega}{2\pi T_0}\right)$ and $\psi^{(1)}(x)$ is the trigamma function.
Let us also introduce the dimensionful ($g$) and dimensionless ($\lam$) electron-phonon coupling constants,
\begin{align}
g^2=2\int d\omega \, \omega\alpha^2 F(\omega)=\sum_k \lam_k\omega_k^2,\label{g}\\
 \lam=2\int d\omega \frac{\alpha^2 F(\omega)}{\omega}=\sum_k\lam_k.
\end{align}

We first analyze the strong coupling limit $\lam\to\infty$, where the instability is most pronounced. Another attractive feature of this limit is its universality: since $\lam\to\infty$ is equivalent to taking all phonon frequencies to zero, the results are independent of the phonon dispersion~\cite{combescot}.  Finite differences turn into  derivatives, e.g., $f_+-f_-\to 2f'\omega$, and the r.h.s. of \eref{ke} becomes $\pi g^2 (T_0 f''-2f f' +f')$.
It is convenient to work in energy units $g=1$  and to rescale the time $\pi t\to t$. In these units, $T_c=0.1827$ in the limit $\lam\to\infty$~\cite{allendynes}.  For small deviations from equilibrium and $E\gg T$, the $f'\dot \Sigma$ term on the l.h.s. of \eref{ke} is negligible and $1-\Sigma'\approx Z_0^{-1}$ to the leading order in the deviation. \eref{ke} therefore takes the form 
\beg
Z_0^{-1}\dot f=T_0 f''-2ff'+f'.
\label{burger1}
\en
In terms of new variables $u=2f-1$ and $y=E-\Sigma[f_0]$, this is the well-known Burgers' equation~\cite{burgers,Whitham} with energy (position) dependent viscosity $T_0 Z_0^{-1}(E)$,
\beg
\dot u = \frac{d}{dy}\left(T_0 Z_0^{-1} \frac{du}{dy}\right) - u \frac{du}{dy}.
\label{burger2}
\en
 The $2ff'$ term is similarly negligible for $E\gg T$ and therefore $u \frac{\partial u}{\partial y}\approx - \frac{\partial u}{\partial y}$. With this replacement, \eref{burger2} becomes the 1D convection-diffusion equation with position-dependent diffusion coefficient $T_0 Z_0^{-1}(E)$. The expression~\re{zfull} for $Z^{-1}_0(E)$  simplifies in the strong coupling limit as well~\cite{app}.
Plotting  it for various $T_0$, we observe that for $T_0\le 1.63T_c$  there is a region where it dips below the horizontal axis. In particular, $Z_0(E)< 0$ for $0.40 <E< 1.16$ at $T_0=T_c$. 

The existence of a region of negative viscosity in Burger's equation~\re{burger2} or, equivalently, negative diffusion coefficient in the diffusion-convection equation indicates  an instability. In the region where $Z_0(E)<0$, deviations of the electron distribution function $f$ from equilibrium grow and aggregate instead of decreasing and spreading out~\cite{Pelinovskii,karpov}. Flipping the sign of $Z_0^{-1}$   in \eref{burger2} is equivalent to reversing the arrow of time. As a result, the system moves away from rather than toward the equilibrium. Another way to  see this  is to notice that, according to \eref{burger1},  $\dot f>0$ at the maxima of $f(E,t)$ [$f'=0$ and $f''<0$] and
$\dot f<0$ at the minima when $Z_0<0$. Opposite to deviations from  stable equilibria,    maxima of $f(E,t)$ grow and minima decrease. Straightforward numerical simulations of the full nonlinear kinetic equation~\re{ke} confirm this behavior (see Fig.~\ref{FigInstability}), i.e., the system is unstable whenever 
\beg
\min_E Z^{-1}_0(E)<0.
\label{min}
\en
Note also that neglecting the $-2ff'$ term in \eref{burger1} and looking for a solution  of the form $f\propto e^{-\theta t +i kE}$, we find
\beg
\gamma(E)\equiv\Re[ \theta(E)]=  T_0 Z_0(E) k^2. 
\label{gamma}
\en
Although this solution is good only at small times and for $E\gg T_0$, it reproduces the main features of the exact solution of the full kinetic equation~\re{ke}. In particular, we see that the solution grows in regions where $Z_0<0$. The growth rate is largest for large $k$, indicating a highly oscillatory behavior, and diverges at the two points where $Z^{-1}_0(E)=0$, cf. Fig.~\ref{FigInstability}.

\begin{figure}[h!]
\includegraphics[width=0.46\textwidth]{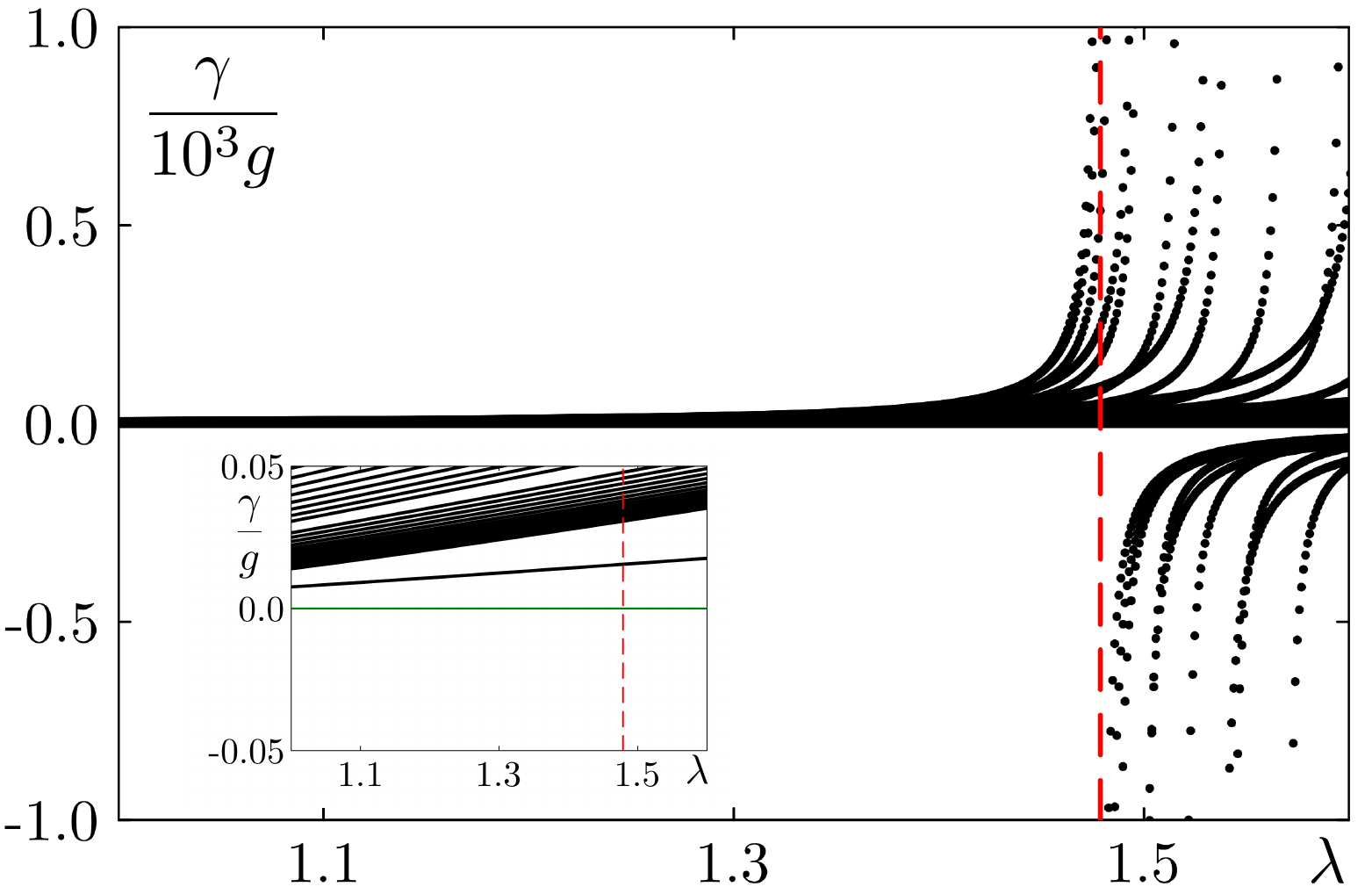}
\caption{Results of the linear stability analysis: generalized eigenvalues $\gamma$ vs the electron-phonon coupling $\lam$ at temperature $T_0=0.20g$. The metal is  unstable when at least one $\gamma<0$. At this temperature, the instability takes place at $\lam_c=1.4788$.  The inset shows the part of the spectrum with smallest $|\gamma|$. Note the polelike divergencies in $\gamma(\lam)$. The first of them is at $\lam=\lam_c$ when the inverse quasiparticle weight first touches the horizontal axis, i.e.,  $\min_E Z^{-1}_0(E)=0$, see the main text and~\cite{app} for more details.  Phonon frequencies and couplings $\lam_k$ are the same as in Fig.~2 of \cite{app}.}
\label{FigEigSpec}
\end{figure}

To investigate this instability at finite $\lam$ and for various phonon spectra, we perform linear stability analysis of the kinetic equation~\re{ke}. Linearizing this equation in small deviations from equilibrium and  substituting (Sec.~79 of \cite{phys_kin})
\beg
\delta f=f-f_0=-f_0' \varphi,
\label{subst}
\en 
we obtain
\beg
A\cdot \dot\varphi=-M\cdot \varphi,
\label{lin}
\en
where $A\cdot \dot\varphi=\int d\tilde E A_{\tilde E E} \dot\varphi(\tilde E)$ and similarly for $M\cdot \varphi$. The metallic state is stable when all separable solutions of \eref{lin} decay exponentially, 
$\varphi\propto e^{-\gamma t}$ with $\gamma>0$, i.e., when all eigenvalues $\gamma$ of the generalized eigenvalue equation
\beg
\gamma A\cdot \varphi=M\cdot \varphi
\label{eg}
\en 
are positive. Since $A$ and $M$ are real symmetric and $M$ is positively defined~\cite{app}, $\gamma$ are real. It is straightforward to see that at weak coupling, $\lam_k\to 0$, $A$ is  positively defined as well~\cite{app}  and, therefore, all $\gamma>0$. As we increase the coupling, one of the eigenvalues diverges at $\lam=\lam_c$ (when $A$ becomes singular) and becomes negative for $\lam>\lam_c$  (Fig.~\ref{FigEigSpec}). The same behavior is seen in \eref{gamma}, and $\lam_c$  coincides with $\lam$ for which   $\lam_c$ obtained from $\min_E Z^{-1}_0(E)=0$ up to a relative error of order of $10^{-4}$~\cite{app}. Note that   initial conditions are irrelevant in linear stability analysis; instability arises if a single normal mode is unstable.

It is important to mention that the stability condition~\re{min} underestimates   $\lam_c$ due to the quasiclassical nature of the kinetic equation~\re{ke}, which ignores the quantum uncertainty   in the electron energy~\cite{rammer}.  The length $\delta E$ of the interval of energies where $Z_0(E)<0$ needs to be comparable to  the energy uncertainty $\Delta E$ to ensure the instability. Consider, for example,  Einstein phonons, $\alpha^2F(\omega)=\frac{\lam\Omega}{2}\delta(\omega-\Omega),$ at $T=0.183g >T_c$. Using \eref{zfull},
we find that $Z_0(E)$ touches the horizontal axis at $E_{\min}=1.44\Omega$ and $\lam_c=1.27$. Since all relevant energies are of  the order of $\Omega$, the uncertainty $\Delta E\sim \Omega$. However,   $\delta E\ge \Omega$  only for $\lam\ge3.40$. On the other hand, $\delta E>\Delta E$ already at $T=T_c$ in the strong coupling regime~\cite{app}.

The instability of the metallic state  we discovered is an absolute instability (as opposed to metastability). Instabilities of this type are often accompanied by the violation of fundamental thermodynamic inequalities. Indeed, we mentioned earlier that the electronic specific heat turns negative at sufficiently large $\lam$.  Even though electrons are strongly coupled to the phonons and the total specific heat is positive~\cite{breakdown}, the fact that electrons   retain many features of an independent system~\cite{prange1}   does suggest that $C_\mathrm{el}>0$ is required for stability as usual. In particular, the rate of change of the total energy of the electron subsystem and, as a consequence, $C_\mathrm{el}$ depend only on the electron distribution $f(E, t)$ (and not on the phonon distribution). Moreover, the r.h.s. of the kinetic equation~\re{ke} has the same form for any electron-phonon coupling. This suggests that the heat flows from hot to cold as usual. If the electrons are initially hotter and $C_\mathrm{el}<0$, their temperature will only increase and the system will never equilibrate. 

To see this explicitly, i.e., to prove that $C_\mathrm{el}<0$ is a \textsl{sufficient} condition for the instability, it is helpful to incorporate the electron-electron collisions into the kinetic equation~\re{ke},
\beg
 (1-\Sigma') \dot f +f' \dot\Sigma =  a_\mathrm{ep}I_\mathrm{ep}+a_\mathrm{ee}I_\mathrm{ee},
 \label{ke1}
  \en 
  where $I_\mathrm{ee}$ is the electron-electron collision integral. This corresponds to  $M=a_\mathrm{ep}M^\mathrm{ep}+a_\mathrm{ee}M^\mathrm{ee}$ in the linearized kinetic equation~\re{lin}, where $M^\mathrm{ee}$ is the linearized $I_\mathrm{ee}$. We weighted the collision integrals with positive coefficients $a_\mathrm{ep}$ and $a_\mathrm{ee}$ for the sake of the argument; in the actual dynamics $a_\mathrm{ep}=a_\mathrm{ee}=1$. 
Suppose the metal is stable with respect to the electron-phonon interaction ($a_\mathrm{ep}=1$ and $a_\mathrm{ee}=0$), i.e.,
all eigenvalues $\gamma>0$ when $M=M^\mathrm{ep}$. Then, \eref{eg} implies that   $A$ is positively defined. Since any linear combination
$M=a_\mathrm{ep}M^\mathrm{ep}+a_\mathrm{ee}M^\mathrm{ee}$ of positively defined real-symmetric matrices $M^\mathrm{ep}$ and $M^\mathrm{ee}$~\cite{app} with positive coefficients is positively defined and given that $A$ is positively defined as well, it follows that
$\gamma>0$ also for $M=a_\mathrm{ep}M^\mathrm{ep}+a_\mathrm{ee}M^\mathrm{ee}$ with arbitrary positive $a_\mathrm{ep}$ and $a_\mathrm{ee}$. In other words, if the metal is stable with respect to strong electron-phonon interaction, \eref{ke1} must also be stable for arbitrary $a_\mathrm{ep}>0$ and $a_\mathrm{ee}>0$. 

Let $a_\mathrm{ee}\gg a_\mathrm{ep}>0$. Then, electron-electron collisions are much faster and, therefore,    $f(E,t)$ quickly equilibrates, $f(E,t)=f_0[T(t)]$, at a temperature $T(t)$ corresponding to the instantaneous total energy of the electronic subsystem. 
To see this, note that $I_\mathrm{ee}$ vanishes  for  $f(E,t)=f_0(T)$ for any $T$ and electron-electron collisions conserve the total electron energy density $\varepsilon$. Indeed, multiplying \eref{ke} by $E$ and integrating, we find~\cite{prange1,app}
$$
 \dot\varepsilon=2\nu_0\int\!\! dE\, E\left[(1-\Sigma') \dot f +f' \dot\Sigma\right] =2\nu_0a_\mathrm{ep}\!\! \int \!\!  dE\, EI_\mathrm{ep},
 $$
 since $ \int \!\!  dE\, EI_\mathrm{ee}=0$. Here $\nu_0$ is the density of states at the Fermi level  per  spin projection. Further, substituting $f(E,t)=f_0[T(t)]$ into the last equation, we obtain~\cite{app}  
\beg
C_\mathrm{el}\dot T=4\nu_0 a_\mathrm{ep}\!\! \int_0^\infty \!\!\! d\omega\alpha^2 F(\omega)\omega^2\left[ N_0(T_0)-N_0(T)\right].
\label{inst}
\en
By definition, a kinetically stable system must be stable for any small deviation from the equilibrium, so it is sufficient to demonstrate instability for a particular  initial condition. Suppose $f(E, t=0)=f_0(T_\mathrm{in})$ with $T_\mathrm{in}$ slightly  above $T_0$. Since $\alpha^2F\ge0$ and
$N_0(T)$ is a monotonically increasing function of $T$, the r.h.s. of \eref{inst} is negative for any $T>T_0$.  If $C_\mathrm{el}<0$, $T(t)$ will grow exponentially proving that our assumption that the metal is stable is incorrect. 

We emphasize that our proof does \textsl{not} assume that electrons \textsl{actually} equilibrate among themselves quicker than with the lattice. This does happen under certain circumstances justifying the so-called two-temperature model, where electrons are in quasiequilibrium at their own temperature $T(t)$, which slowly relaxes to the lattice temperature~\cite{kaganov,anisimov,optical1,optical2,Komashko,twoTreview,optical3}.   This model is frequently employed in the studies of thermal relaxation of photoexcited   metals and operates with equations similar to \eref{inst}. In contrast, in our proof \eref{inst} emerges as a consequence of the assumption of stability and does not necessarily describe the actual dynamics. In other words, the statement is that \eref{inst} must be stable provided that the metal is linearly stable with respect to the electron-phonon interaction. This is true regardless of whether or not electrons equilibrate among themselves quicker than with the lattice in  the course of their actual dynamics. The latter is described by \eref{ke1} with $a_\mathrm{ee}=a_\mathrm{ep}=1$ and not \eref{inst}. Also, note that we need not worry about the energy uncertainty in this case, since $C_\mathrm{el}$ is an integral over all energies.

\begin{figure}
\includegraphics[width=0.49\textwidth]{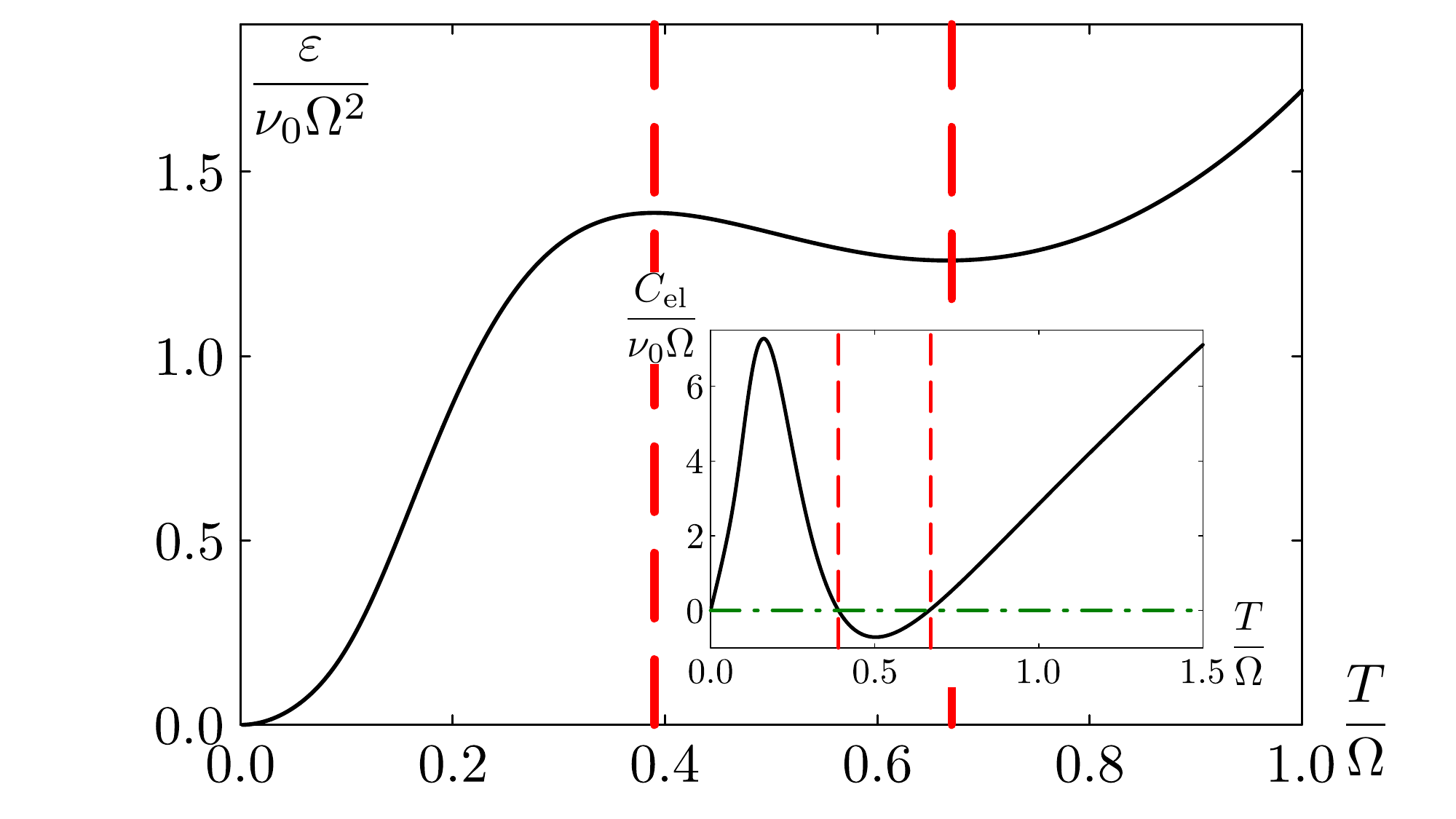}
\caption{  Electronic internal energy density  $ \varepsilon$  vs $\frac{T}{\Omega}$ for Einstein phonons with $\lambda = 4.5$ (inset: plot of the electronic specific heat $C_\mathrm{el}$). The metal is absolutely unstable in the temperature range from $T_- = 0.39\Omega$ to $T_+ = 0.67\Omega$ (vertical dashed red lines). For $T_-<T<T_+$, the slope of the $ \varepsilon(T)$ is negative, i.e.,
$C_\mathrm{el}<0$.  Cf. isotherms $p(v)$ of a van der Waals gas that similarly feature regions, $v_-<v<v_+$ of absolute instability where the compressibility   $\beta=-\frac{\partial p}{\partial v}<0$. The system therefore cannot exist in a purely metallic state  for $T_-<T<T_+$ similarly to how the van der Waals gas  cannot exist in a purely gaseous or purely liquid state for $v_-<v<v_+$.
}
\label{Fig1stOrderPT}
\end{figure}

The  above instability of metals is  analogous to that of  van der Waals gases. Recall that below a certain critical temperature an isotherm $p(v)$ of this gas shows a  segment $v_-<v <v_+$with positive slope, i.e., negative compressibility $\beta=-\frac{\partial p}{\partial v}<0$~\cite{epstein,callen}.  Similarly, the curve $\varepsilon(T)$---electron internal energy density as a function of temperature---shows a segment $T_-<T<T_+$ with negative slope [$C_\mathrm{el}(T)<0$], see Fig.~\ref{Fig1stOrderPT}. In both cases, this signals a discontinuous (first order) phase transition, but does not by itself determine the nature of the new phase.  From complementary considerations, we know that it is a gas-liquid transition for the gas, and a structural transition to an insulator or another metal with weaker electron-phonon interaction for the metal~\cite{breakdown}. 

In contrast to metastable states, such as an overheated liquid, states with $\beta<0$ or $C_\mathrm{el}<0$ are absolutely unstable, i.e., unstable already to infinitesimal deviations from the equilibrium.  Consider a gas with $\beta<0$ confined to a vertical cylinder by a piston with a load that  balances the pressure of the gas~\cite{epstein}. Suppose we slightly decrease  the weight of the load. The gas will then expand, but the pressure will grow, and the piston will continue to accelerate upward until  it reaches the point $v=v_+$. If we increase the weight of the load, the piston will accelerate downward until $v=v_-$. Similarly, if $C_\mathrm{el}<0$ and the electron temperature $T$ is initially slightly above (below) the lattice temperature, it will grow (decrease) until $T=T_+$ ($T=T_-$).  

 It is important to bear in mind that the kinetic instability  we identify is an inherently  nonequilibrium phenomenon, and its proof does not rely on  ME theory.
The kinetic equation becomes unstable    \textsl{before} $C_\mathrm{el}$ becomes negative~\cite{example}. In other words, the  instability arises before the equilibrium ME theory is able to detect it. Further underscoring the  fundamental difference from the  P\slash CDW transition is that the kinetic instability can proceed along entirely different routes.  A representative  example is provided by hydrides such as H$_3$S, LaH$_{10}$, and YH$_9$,  where increasing  $\lam$ through decompression results in hydrogen diffusion and eventual formation of new  compounds rather than a P\slash CDW transition~\cite{kong,chen,drozdov1,minkov1,chen3,bi1}. This process provides a natural experimental setting to probe the onset of the kinetic instability discussed here. 
Although the initial dynamics of the instability is uniform, spatial structure may emerge later due to secondary instabilities involving couplings across length scales~\cite{app};  a detailed analysis of this late-time behavior is beyond the scope of the present work.

\vspace{3mm}

\textsl{Acknowledgements}---We thank A. E. Allahverdyan, R. Boyack, A. V. Chubukov,   I. V. Gornyi, J. T. Heath, A. Kamenev, M. K.-H. Kiessling, G. Kotliar, D. V. Semenok, J. Schmalian, and D. Vanderbilt for stimulating discussions. A.P. acknowledges the financial support from the Institute for Basic Science (IBS) in the Republic of Korea through Project No. IBS-R024-D1.

\vspace{3mm} 

\textsl{Data availability}---The source code required to reproduce figures in this Letter and Supplemental Material  is openly available at~\cite{zenodo}.

\clearpage
\onecolumngrid
\begin{center}
 \textbf{\Large Supplemental Material}
\end{center}

\vspace{7mm}

\twocolumngrid

\setcounter{equation}{0}
\setcounter{figure}{0}
\setcounter{table}{0}
\setcounter{section}{0}
\makeatletter
\renewcommand{\theequation}{S\arabic{equation}}
\renewcommand{\thefigure}{S\arabic{figure}}
\renewcommand{\thetable}{S\arabic{table}}
\renewcommand{\bibnumfmt}[1]{[S#1]}
\renewcommand{\citenumfont}[1]{S#1}
\renewcommand{\thesection}{S\arabic{section}}
\renewcommand{\theHtable}{S\thetable}
\renewcommand{\theHfigure}{S\thefigure}
\renewcommand{\theHequation}{S\theequation}
\section{Electronic specific heat}
\label{S1}

We define the grand canonical partition function for the electron subsystem as (in our units $k_B=\hbar=1$)
\beg
\mathcal{Z}_\mathrm{el}=\frac{ \mathcal{Z} }{ \mathcal{Z}_\mathrm{ph} }=\frac{ \Tr\,   e^{ -(\hat H-\mu\hat N_\mathrm{el})/T}   }{ \Tr_\mathrm{ph}\, e^{-\hat H_\mathrm{ph}/T}  },
\label{partition}
\en
where $ \mathcal{Z}$ is the full partition function of the electron-phonon system and $\mathcal{Z}_\mathrm{ph}$ is the partition function of noninteracting phonons. This definition for the partition function of an interacting subsystem or, equivalently, of a system coupled to a  bath (recall that phonons have a much larger heat capacity and effectively serve as a thermal bath for the electrons) arises naturally within the effective action approach~\cite{feynman1,feynman2,Caldeira,Ingold,weiss}. As discussed in the introductory part of the main text, one should not renormalize the phonon frequencies within the ME theory but borrow them from  ab initio calculations or experiment instead. For this reason, we take $\hat H_\mathrm{ph}$ in \eref{partition} to be the noninteracting phonon Hamiltonian but with the physical rather than bare phonon energies.

Accordingly, we define the electronic grand (Landau) potential $\mathtt{ f}_\mathrm{el}$ per lattice site and the electronic specific heat $C_\mathrm{el}$ through
\begin{align}
e^{-N \mathtt{f}_\mathrm{el}/T }=\mathcal{Z}_\mathrm{el},\\
C_\mathrm{el}=-T \frac{\partial^2 \mathtt{f}_\mathrm{el} }{\partial T^2},
\label{Cdef}
\end{align}
where $N$ is the number of lattice sites. It follows from these standard definitions that the total grand potential $\mathtt{f}$ and total specific heat $C$ are
\beg
\mathtt{f}=\mathtt{f}_\mathrm{el}+\mathtt{f}_\mathrm{ph},\quad C=C_\mathrm{el}+C_\mathrm{ph},
\en
where $\mathtt{f}_\mathrm{ph}$ and $C_\mathrm{ph}$ are the   grand potential and specific heat of noninteracting phonons, respectively.

Since the specific heat of a metal $C$ and the specific heat of noninteracting phonons $C_\mathrm{ph}$ are both  well-defined and finite,  $C_\mathrm{el}=C- C_\mathrm{ph}$ is  free of any divergences when calculated correctly. Divergences   discussed in~\cite{spinchain_sup,breakdown_sup,largegamma,zhang1,zhang2} are merely artifacts of an inaccurate expression for $\mathtt{f}_\mathrm{el}$ on the Matsubara axis. We employed the zeta-function regularization procedure to obtain the correct answer~\cite{spinchain_sup,breakdown_sup,largegamma} (see below), while~\cite{zhang1} performed a cutoff regularization. In contrast, Prange and Kadanoff obtain the following answer on the real axis~\cite{prange1_sup}, which is divergence free  from the beginning  without any regularization (see also \cite{grimvall}):
\beg
\begin{split}
C_\mathrm{el}=  2\nu_0\!\! \int \!\!  dE\, E\left\{ \left[ 1-\frac{\partial\Sigma}{\partial E}\right]\frac{\partial f_0}{\partial T}+ \frac{\partial\Sigma}{\partial T} \frac{\partial f_0}{\partial E}\right\}\\
= \frac{2}{3}\pi^2\nu_0 T+
\frac{2\nu_0}{T} \frac{\lam\Omega}{2} 
 \dashint \!  dE d\tilde E   \frac{\partial f_0(E)}{\partial E}  \frac{\partial f_0(\tilde E)}{\partial \tilde E}\\
 \times \frac{(E-\tilde E)^2}{E-\tilde E+\Omega},
\end{split}
\label{cpr}
\en
where $\Sigma$  is the equilibrium self-energy given by \eref{se1} with $f\to f_0$ and $f_0$ is the equilibrium  Fermi distribution function.  For simplicity, we specialized the second and third lines of \eref{cpr} to Einstein phonons with energy $\Omega$. To write down the answer for a general Eliashberg function, simply replace 
$\frac{\lam\Omega}{2}\to \int_0^\infty \!\! d\omega \alpha^2F(\omega)$ and $\Omega\to\omega$ in the denominator in the third line.  Prange and Kadanoff derived \eref{cpr} from the kinetic equation~\re{ke} by considering the change in the energy density $\varepsilon$ of the electronic subsystem in response to  a gradual change of temperature and using 
\beg
\dot \varepsilon =\int\! dE E\{[1-\Sigma']\dot f +\dot\Sigma f'\}.
\label{rate_sup}
\en
 Note that there is a typo in~\cite{prange1_sup}---the term in the square brackets in \eref{cpr} should have the same sign as in the kinetic equation, i.e., minus rather than  plus. Note also that with the help of  \esref{cpr} and \re{rate_sup}  and the chain rule, we immediately find the rate of change of the energy density for $f(E,t)=f_0(T(t))$ in terms of $C_\mathrm{el}$,
\beg
 \dot \varepsilon =C_\mathrm{el}\dot T.
\en
Prange and Kadanoff used this result to derive the expression~\re{cpr} for $C_\mathrm{el}$ and we used it to obtain \eref{inst}.

Lee and Rainer showed~\cite{lee1} that  \eref{cpr}  corresponds to the following expression for the  Landau potential on the Matsubara axis:
\beg
 \mathtt{f}_\mathrm{el}=-\frac{\pi^2\nu_0 T^2}{3} -\pi^2\nu_0 T^2 \lam\Omega^2 \sum_{nm} \frac{ \mathrm{sgn}(\omega_n\omega_m)-1}{(\omega_n-\omega_m)^2 +\Omega^2},
 \label{Lreg}
\en
where $\omega_n=\pi T(2n+1)$ are the fermionic Matsubara frequencies, i.e.,  \eref{cpr} obtains  by differentiating \eref{Lreg} twice with respect to $T$, see  \eref{Cdef}. On the other hand, a straightforward evaluation of  the continuous time  path integral  produces a divergent result~\cite{spinchain_sup,breakdown_sup,largegamma}
\beg
 \mathtt{f}_\mathrm{el}=-2\pi\nu_0 T \sum_n| \omega_n|-\pi^2\nu_0 T^2 \lam\Omega^2 \sum_{nm} \frac{ \mathrm{sgn}(\omega_n\omega_m)}{(\omega_n-\omega_m)^2 +\Omega^2}.
 \label{Lunreg}
\en
Indeed, it is well known that the evaluation of  path integrals without proper time slicing and a precise definition of the integration measure  often results in badly divergent expressions~\cite{kleinert,negele,Atland}. This is  already so for, e.g., free particles or a harmonic oscillator. The same divergences as in \eref{Lunreg} arise in the Luttinger-Ward free-energy functional after integrating over the single-electron energy~\cite{lee1}. 
The way out is  either to do the  path integral correctly or to properly regularize the divergencies. Both approaches result in \eref{Lreg}. 

Clearly, not any regularization method will work. In particular, the cutoff regularization of \eref{Lunreg} does not produce the right answer, in particular because the cutoff dependence does not cancel out. As Lee and Rainer point out~\cite{lee1}: ``Obviously, the sum in (1)  diverges, and cannot be regularized in a simple way, for instance by a frequency cut off.'' Here (1) refers to the low-temperature limit of our \eref{Lunreg}.

A well-known technique for dealing with divergent functional determinants is the zeta-function (also known as analytic) 
regularization~\cite{kleinert,Hawking,bytsenko,encyclop}, which does lead to the correct result~\cite{spinchain_sup,breakdown_sup,largegamma}. Specifically, we manipulate the first term in \eref{Lunreg} as
\begin{equation}
\nonumber
\begin{split}
&-2\pi\nu_0 T \sum_n| \omega_n|=-4\pi^2 T^2\nu_0\sum_{n=-\infty}^\infty \left| n+\frac{1}{2}\right|\\
& =  -8\pi^2 T^2\nu_0\sum_{n=0}^\infty \frac{1}{\left( n+ \frac{1}{2}\right)^{-1}}
\to -8\pi^2 T^2\nu_0\zeta\left(-1, \frac{1}{2}\right)\\
&\qquad \qquad \qquad \qquad \qquad \qquad \quad=-\frac{\pi^2\nu_0 T^2}{3},
\end{split}
\end{equation}
where $\zeta(s, a)=\sum_{n=0}^\infty \frac{1}{(n+a)^s}$ is the Hurwitz zeta function and we used $\zeta\left(-1, \frac{1}{2}\right)=\frac{1}{24}$. Another way to arrive at this result is to use the Poisson summation formula or convert the Matsubara sum to a contour integral with a suitable convergence factor~\cite{Piers}. 

In the second term in  \eref{Lunreg}, we add and subtract 1 in the numerator. This results in a convergent expression [the second term on the RHS of \eref{Lreg}]  plus a divergent extra term,
\beg
-\pi^2\nu_0 T^2 \lam\Omega^2 \sum_{nm} \frac{1}{(\omega_n-\omega_m)^2 +\Omega^2}.
\label{const}
\en
Applying the zeta-function regularization prescription to this term, we find
\beg
\nonumber
\begin{split}
\pi^2 T^2 \sum_{nm} \frac{1}{(\omega_n-\omega_m)^2 +\Omega^2}= \frac{1}{4} \sum_{nm} \frac{1}{(n-m)^2+\left(\frac{\Omega}{2\pi T}\right)^2 }\\
= \frac{1}{4} \sum_{L} \sum_{l} \frac{1}{l^2+\left(\frac{\Omega}{2\pi T}\right)^2 }
= \sum_{L=-\infty}^\infty \mathrm{const}\\
=\mathrm{const} \left[ 1+ 2 \sum_{L=1}^\infty \frac{1}{L^0}\right]\to \mathrm{const}[1+2\zeta(0)]=0.
\end{split}
\en
Here $L=n+m$, $l=n-m$ and we used $\zeta(0)=-\frac{1}{2}$.  In fact, divergent Matsubara sums of the from $\sum_n \mathrm{const}$ that regularize to zero are ubiquitous  when evaluating functional integrals without  proper time slicing, see, for example, p.~164 in~\cite{kleinert}. We see that the zeta-function regularization does indeed produce the correct answer~\re{Lreg}.

Lee and Rainer further evaluated \eref{cpr} explicitly in terms of the digamma function $\psi(z)$~\cite{lee1},
\beg
C_\mathrm{el}=  \frac{2\pi^2 \nu_0 T}{3}\left[ 1- \lam g\left(\frac{\Omega}{2\pi T}\right)\right],
\label{cn}
\en
where
\beg
g(x)=6x^2+12 x^3 \mbox{Im}[\psi^{(1)}(i x)]+6x^4\mbox{Re}[\psi^{(2)}(ix)],
\label{g}
\en
where $\psi^{(1)}(z)$ and $\psi^{(2)}(z)$ are the first and second derivatives of the digamma function [$\psi^{(1)}(z)$ is also known as the trigamma function].
The digamma function  enters through the  equilibrium normal state self-energy. Indeed, evaluating $\Sigma[f]$ according to its definition~\re{se1}  for $f=f_0(T)=[e^{\frac{E}{T}}+1]^{-1}$, we find 
\beg
 \Sigma= \frac{\lam\Omega}{2}\Re(\psi_+-\psi_-),
 \en
where $\psi_\pm=\psi\left(\frac{1}{2}+\frac{iE\pm i\Omega}{2\pi T}\right)$. Incidentally,  we used this result to derive the equilibrium inverse quasiparticle weight function~\re{zfull}.

Let us  estimate $\frac{|C_\mathrm{el}|}{C_\mathrm{ph}}$ at relevant temperatures for $\lam\gtrsim 1$. We will employ the 3D Debye model   for definiteness, even though a similar result holds in 2D and for Einstein phonons.  We have $T_c > 0.10 - 0.18 \Omega$ for $\lam\gtrsim 1$~\cite{allendynes_sup,mitrovic_sup,carbotte_sup}, and we used $T=0.18\Omega$ as our lowest temperature in the main text. The phonon  heat capacity at $T=0.18\Omega$ is $0.92 N$, where $N$ is the number of the ions~\cite{kittel}. According to \eref{cn}, $C_\mathrm{el} \sim
\frac{2\pi^2 NT}{3E_F}$ assuming one conduction electron per ion. We conclude that indeed $|C_\mathrm{el}|/C_\mathrm{ph} \sim \Omega/E_F$ as stated in the main text.

Recently, Zhang \textsl{et al.} presented~\cite{zhang2_sup} one more calculation of $C_\mathrm{el}$ (for Einstein phonons in 2D)~\cite{note}, which is  overall similar  to the one by Lee and Rainer~\cite{lee1}. Just as~\cite{lee1} it starts with the  Luttinger-Ward functional and obtains the same well-known answer for  $C_\mathrm{el}$ (see, e.g.,~\cite{breakdown_sup,dolgov,golubov}) that Lee and Rainer originally derived in~\cite{lee1}, though  the intermediate steps are not the same. In particular, an interesting   feature of~\cite{zhang2_sup}   is  that the ``counter term''~\re{const} emerges naturally, while \cite{lee1} merely suggests  that it must arise from the phonon part of the Luttinger-Ward functional without providing further details.  

In addition, Zhang \textsl{et al.}   conjecture~\cite{zhang2_sup}  that the normal state is stable even when $C_\mathrm{el}<0$ just because  $|C_\mathrm{el}|$ is parametrically smaller than the phonon specific heat $C_\mathrm{ph}$ and the total specific heat  $C_\mathrm{el}+C_\mathrm{ph}>0$.  However, the smallness of $C_\mathrm{el}$ is not a valid argument in favor of stability. As a matter of fact, a system with a small negative heat capacity connected to another system (thermal bath) with positive and much larger in magnitude  heat capacity (thus resulting in a positive total specific heat) is a textbook example of an absolute instability of this type~\cite{sp1,sp2,sp3}. The situation is more complicated in our case since the  interaction between the electron and phonon systems is not weak. Due to this, a more detailed analysis of the deviations from the the thermal equilibrium between the electrons and phonons, such as the one we performed  in the main text, is necessary  to  prove or disprove the instability decisively.

 In practice, the stability condition $C_\mathrm{el} > 0$ can be directly tested for any metal with a known Eliashberg function. The specific heat for a general $\alpha^2F(\omega)$ takes the form~\cite{lee1}
\beg
C_\mathrm{el}=  \frac{2\pi^2 \nu_0 T}{3} \left[ 1-  \int_0^\infty g\left( \frac{\omega}{2\pi T}\right)\frac{2  \alpha^2 F(\omega)}{\omega} d\omega\right]\!,
\label{C}
\en
with $g(x)$ given by \eref{g}. 
The requirement $C_\mathrm{el} > 0$  thus imposes a quantitative constraint on the Eliashberg function $\alpha^2 F(\omega)$.

\section{Details of the linear analysis}
\label{linear_sup}

Recall that the linear kinetic stability analysis reduces to solving the generalized eigenvalue equation~\re{eg},
\beg
\gamma \int \!\! d\tilde E\, A_{\tilde E E}  \varphi_{\tilde E}= \int \!\!  d \tilde E \, M_{\tilde E E}  \varphi_{\tilde E}.
\label{eg_sup}
\en 
The metal is stable when all generalized eigenvalues $\gamma$ are positive and unstable when at least one of them is negative.
The matrix (integration kernel) $A_{\tilde E E} $ reads
\beg
A_{\tilde E E}=-Z^{-1}_0 f'_0 \delta_{\tilde E,E}+\sum_k \frac{\lam_k\omega_k^2 f'_0 \tilde f'_0}{(\tilde E-E)^2 -\omega_k^2}, 
\label{A}
\en
and the matrix $M_{\tilde E E} $ in the case of the electron-phonon collision integral is $M =M^\mathrm{ep}$ and
\beg
M^\mathrm{ep}= \sum_k \lam_k Q^k,
\label{meph}
\en
where (see also \cite{phys_kin_sup})
\beg
\begin{split}
Q^k_{\tilde E E} = \frac{ \omega_k }{T_0} N_0(N_0+1) \left\{ (f_0 - \tilde f_0)  \left[ \delta_{\tilde E,E-\omega_k} - \delta_{\tilde E,E+\omega_k}\right]\right.\\
\left.- \left[f_{0}(E+\omega_k)-f_{0}(E-\omega_k)\right]\delta_{\tilde E, E}\right\}, 
\end{split}
 \label{Q}
\en
 $f_0=f_0(E)$, $\tilde f_0=f_0(\tilde E)$, and $\delta_{X,Y}=\delta(X-Y)$. Whenever not shown, the temperature argument in equilibrium Bose ($N_0$) and Fermi ($f_0$) distributions functions is $T=T_0$. Observe that $Q^k_{\tilde E E}$ and therefore $M^\mathrm{ep}_{\tilde E E}$ are manifestly real symmetric. Keep also in mind that we rescaled the time variable $t\to \pi t$, which is equivalent to $\gamma\to \gamma/\pi$. Without this rescaling, an extra prefactor of $\pi$ will appear in \eref{Q}. 

The matrix $M$ is always positively defined, because the metal is obviously stable when the electron-phonon interaction is infinitesimally weak. Indeed, suppose $\lam_k\ne 0$ only for one value of $k$, i.e.,  only the $k^\mathrm{th}$ phonon mode is present and    $\lam_{k}$ is infinitesimally small. Since $M^\mathrm{ep}=\lam_{k} Q^k$ is already proportional to $\lam_{k}$, the contribution of the self-energy terms to the solution of the generalized eigenvalue equation are higher (second) order in $\lam_{k}$. To determine the solution to the leading (first) order in $\lam_{k}$, we need $A$ only to the zeroth order. To zeroth order,  $Z^{-1}_0=1$ and $A_{\tilde E E}=- f'_0(E) \delta(\tilde E-E)$. This   $A$ is positively defined, since it is a diagonal matrix with a positive diagonal. Given that $\gamma>0$ (the metal is obviously stable at weak coupling) and $\gamma A\cdot\varphi=\lam_{k} Q^k\cdot\varphi$, it is an elementary exercise in linear algebra to show that $Q^{k}$ is positively defined as well. But $Q^k$ is independent of $\lam_k$ and since $M^\mathrm{ep}$ in \eref{meph} is a linear combination of positively defined matrices $Q^k$ with positive coefficients, it is positively defined for \textsl{any} set of positive $\lam_k$, i.e., for any physical Eliashberg function. For the same reason, the matrix $M^\mathrm{ee}$ obtained by linearizing the electron-electron collision integral must also be positively defined.

\begin{figure*}[h!tb]
\includegraphics[width=.96\textwidth]{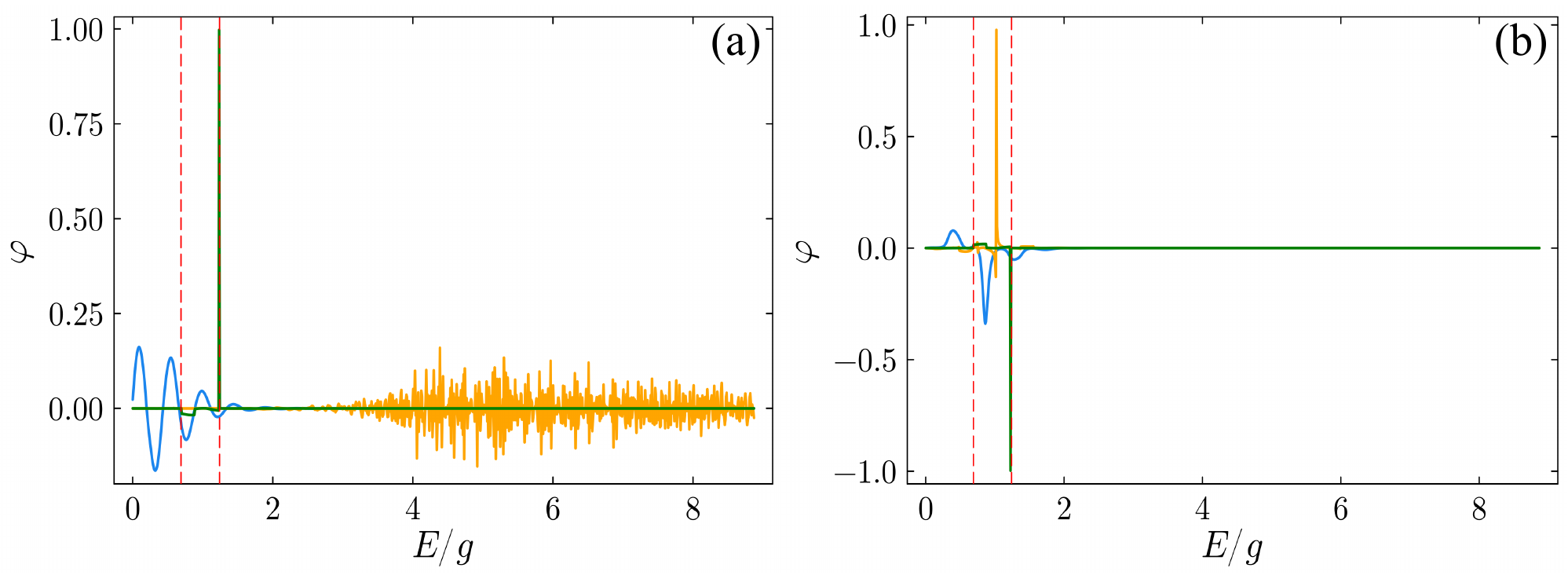}
\caption{Several representative eigenvectors $\varphi_E\equiv \varphi(E)$ of the linearized kinetic equation~\re{eg_sup} in the unstable regime ($\lam=5.0>\lam_c=1.48$). Panel (a) shows three stable eigenvectors and (b) three unstable ones. The corresponding eigenmodes exponentially decay (stable) or exponentially grow (unstable) with  exponents $\gamma$. 
    In both panels, the exponents $\gamma_1, \gamma_2,$ and $\gamma_3$ corresponding to the eigenvectors shown in blue, orange, and green, respectively, satisfy $\left|\gamma_1\right| < \left|\gamma_2\right| < \left|\gamma_3\right|$. In fact, the green (blue) eigenvector in panel (a) corresponds to the largest (smallest) positive eigenvalue, whereas, in panel (b), the same color coding indicates the smallest (largest) negative eigenvalue. Note that unstable eigenvectors are localized in between the two zeros of $Z^{-1}_0(E)$ (shown with vertical dashed red lines), i.e., in the region where $Z^{-1}_0(E)<0$. Phonon frequencies  and couplings to the individual phonon modes  are the same as for Table~\ref{Tab_lam_c}. }
\label{FigEigenvectors}
\end{figure*}

To determine the generalized eigenvalues $\gamma$ numerically, we discretize the energy $E$ in steps of $\epsilon$ and introduce a high energy cutoff $|E|\le \Lambda=L\epsilon$. We also fold the $E$-axis at $E=0$ using $\varphi_{-E}=-\varphi_E$, since, as a consequence of the particle-hole symmetry, $f(-E)=1-f(E)$, and  $\delta f(E)$ is therefore odd in $E$. This results in a matrix equation
\beg
\gamma A \cdot \varphi=M^\mathrm{ep}\cdot \varphi,
\en
where $A$ and $M$ are $L\times L$ real-symmetric matrices. Explicitly,
\beg
\begin{split}
A_{ij}=Z_0^{-1}(E_i) f'_0(E_i)\delta_{ij} - 4\epsilon \sum_k \lam_k  \omega_k^2 \\
\times \frac{  i j f'_0 (E_i) f'_0(E_j)  }{ \left[ (i-j)^2- r_k^2\right] \left[ (i+j)^2- r_k^2\right]}
 \end{split}
\en
and
\beg
\nonumber
\begin{split}
M^\mathrm{ep}_{ij}=\frac{N_0(N_0+1)}{T_0} \sum_k \lam_k \omega_k \Bigl\{  [ f_0(E_{i-r_k}) - f_0(E_{i+r_k}) ] \delta_{ij}\bigr.\\
- [ f_0(E_{i-r_k}) - f_0(E_{i}) ] \delta_{j,i-r_k}\\
 - [ f_0(E_{i}) - f_0(E_{i+r_k}) ] \delta_{j,i+r_k}\\
\Bigl.  + [ 1- f_0(E_{i}) - f_0(E_{j}) ] \delta_{j,r_k-i} \Bigr\},
\end{split}
\en
where $i,j=1,\dots L$, $E_i=i \epsilon$,  and $r_k=\frac{\omega_k}{\epsilon}$ are integers.

We determine $\gamma$ by simultaneously diagonalizing matrices $A$ and $M^\mathrm{ep}$.  At weak coupling, the matrix $A$ is positively defined and therefore all $\gamma>0$. As we increase $\lam$, the smallest eigenvalue of $A$ vanishes at $\lam=\lam_c$. This implies that the largest in magnitude generalized eigenvalue $\gamma\to +\infty$ as $\lam\to\lam_c^-$ and $\gamma\to -\infty$ for $\lam\to\lam_c^+$.  More specifically, $\gamma(\lam)$ has a pole at $\lam=\lam_c$. 
In the main text, we formulated the criterion of instability as [see \eref{min}]
\beg
\min_E Z^{-1}_0(E)<0,
\label{min1}
\en
i.e., $\lam_c$ is the value of $\lam$ for which the graph of $Z^{-1}_0(E)$ touches the $E$-axis. For $\lam$ close to $\lam_c$, $\min_E Z^{-1}_0(E)\propto (\lam_c-\lam)$. Therefore, the largest in magnitude eigenvalue evaluated with the help of the approximate strong coupling expression~\re{gamma}  similarly shows a simple pole at $\lam=\lam_c$ as a function of $\lam$.

While the criterion~\re{min1} holds with high accuracy, it is likely not exact due to the presence of the second term on the RHS of \eref{A}, which originates from the $f'\dot \Sigma$ term in the kinetic equation. However, this term is small due to the smallness of $f'_0(E)$ at the minimum of  $Z^{-1}_0(E)$. As a result, it produces a relative error in $\lam_c$ of order $10^{-4}$ as compared to the instability criterion~\re{min1},   see Table~\ref{Tab_lam_c}. In fact, this error is so small that we cannot confidently discern  whether the $Z^{-1}_0(E)$ criterion is exact or not within the accuracy of our numerics.  

\begin{figure}[h!]
\includegraphics[width=0.43\textwidth]{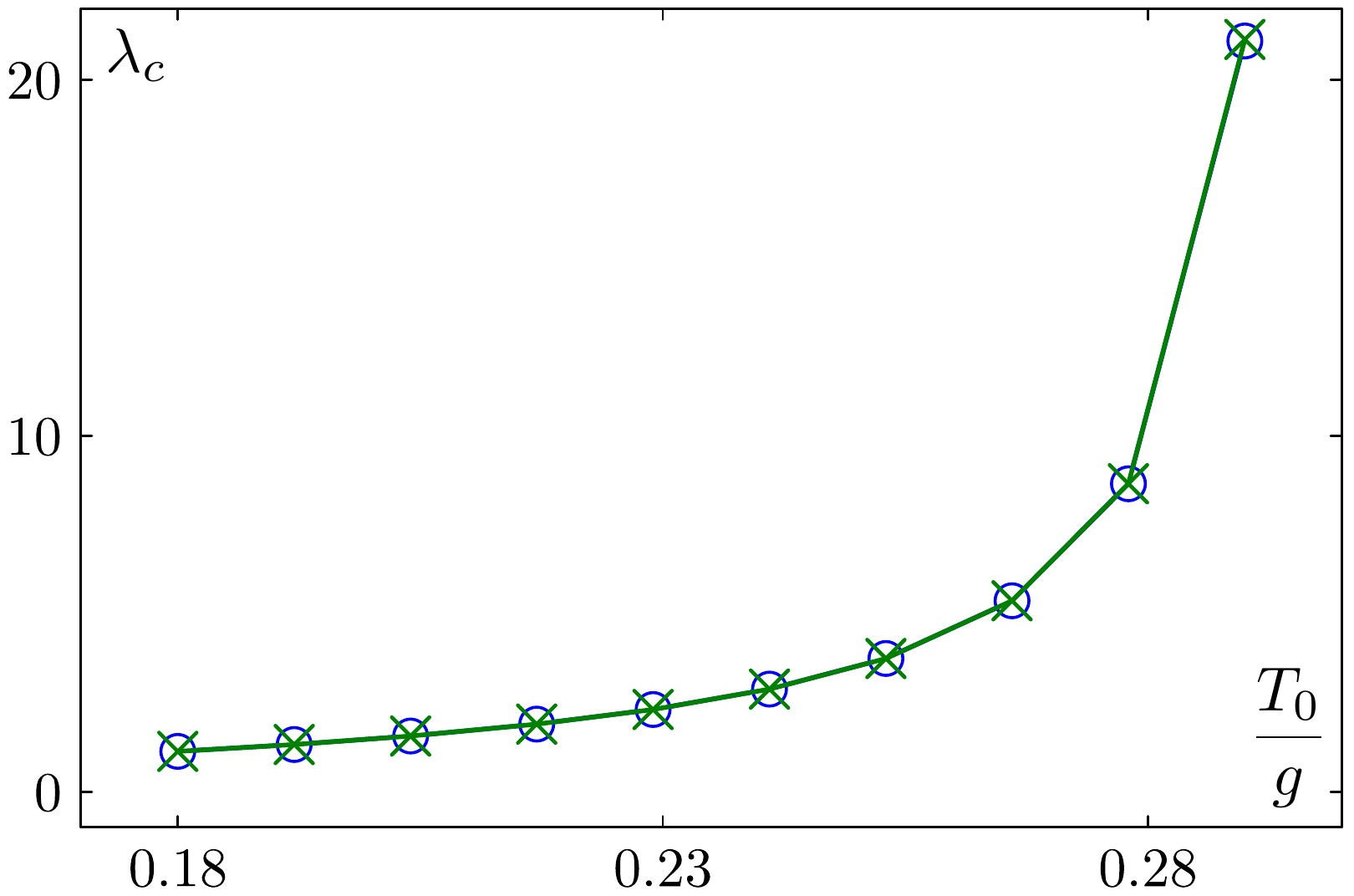}
\caption{Critical values $\lam_c$ of the electron-phonon coupling  vs the equilibrium temperature $T_0$. For $\lam>\lam_c$, the metal  is unstable with respect to small deviations from the thermal equilibrium between the electrons and phonons. Green crosses  obtain from the
instability criterion $\min_E Z^{-1}_0(E)<0$ and blue circles  from the linear stability analysis.  Note that these two methods produce visibly indistinguishable results. $Z_0(E)$ is the quasiparticle weight function. Phonon frequencies are $\omega_k=(39 +k)\epsilon$ with $\lam_k=\frac{\lam}{21}$, where $k=1,2,\dots,21$ and $\epsilon$ is the discrete energy step.  }
\label{lam_c_fig}
\end{figure}

\begin{table}[tbh!]
\centering 
\begin{tabular}{| c | c | c | c |}
\hline
$T_0$ & $\lambda_{c}(\textrm{Linear Analysis})$ &  $\lambda_{c}(\textrm{Minimize } Z^{-1}_0)$ & $\delta\lam_c/\lam_c$\\ [0.5ex]
\hline\hline
0.180 & 1.1335   & 1.1333 & $1.8 \times 10^{-4}$ \\
\hline
0.192 & 1.3262   & 1.3253 & $6.8 \times 10^{-4}$ \\
\hline
0.204 & 1.5644   & 1.5644 &  $< 10^{-4}$ \\
\hline
0.217 & 1.8991  & 1.8980 & $5.8\times 10^{-4}$\\
\hline
0.229 & 2.3096  & 2.3094 & $0.9\times 10^{-4}$ \\
\hline
0.241 & 2.8822  & 2.8820 & $0.7\times 10^{-4}$  \\
\hline
0.253 & 3.7427  & 3.7416 & $2.9\times 10^{-4}$ \\
\hline
0.266 & 5.3639   & 5.3642 & $-0.6\times 10^{-4}$  \\
\hline
0.278 & 8.6508   & 8.6517 & $-1.0\times 10^{-4}$\\
\hline
\end{tabular}
\caption{ Comparison of $\lambda_{c}(T_0)$ values obtained from the full linear stability analysis and from the instability condition $\min_E Z^{-1}_0(E)<0$ [\eref{min1}].  Phonon frequencies $\omega_k$ and couplings to the individual phonon modes $\lam_k$ are the same as in Fig.~\ref{lam_c_fig}.  }
\label{Tab_lam_c}
\end{table}

Let us note that the matrix $A$ is indefinite for $\lam>\lam_c$, and, as a result, some of its eigenvalues vanish in the continuum limit $\epsilon\to 0$. The corresponding $\gamma$ therefore diverge. For discrete energies such divergent eigenvalues occur at a discrete set of values of the coupling $\lam$ (see Fig.~\ref{FigEigSpec}).  To see this, first neglect the small second term on the RHS of \eref{A}. Then, $A$ is diagonal and its eigenvalues are $-Z_0^{-1}(E)f'_0(E)$.
 For $\lam>\lam_c$, $-Z_0^{-1}(E)$ always has two zeroes, which are generally in between of the grid points. The pole-like divergencies seen in Fig.~2  happen when one of the zeros snaps to the grid, i.e., coincides with one of the  $E_i$.  
 
 In Fig.~\ref{FigEigenvectors}, we also show a few representative stable and unstable eigenvectors for $\lam>\lam_c$. All  unstable eigenvectors are localized  between the two zeros of $Z_0^{-1}(E)$ in the region where $Z_0^{-1}(E)<0$. Furthermore, the eigenvectors corresponding to the largest positive (for $\lam<\lam_c$) and the smallest negative (for $\lam>\lam_c$)    $\gamma$  are always localized in the vicinity of the same real root of $Z_0^{-1}(E)$. Note that these eigenvectors  are responsible for the pole-like divergence at $\lam=\lam_c$.

We did not use the explicit form of the electron-electron  collision integral and the corresponding matrix $M^\mathrm{ee}$ in this paper.  The following two properties of the electron-electron  collisions  are important for our proof: (1) they conserve the total energy of the electronic subsystem and (2)  $M^\mathrm{ee}$ is real symmetric. Property (1) reflects energy conservation -- in the absence of interactions with the phonons (with the thermal bath) the total energy of the electrons is conserved. Property (2) manifests itself  thanks to the substitution~\re{subst}, $\delta f=-f_0' \varphi$, see also, e.g., the related discussion in~\cite{LS}.  In addition, we have  verified both these properties   for an  explicit example -- electron-electron collisions in  weakly disordered conductors. The   collision integral in this case reads~\cite{rammer_sup}
\beg
I_\mathrm{ee}(E)=-\frac{1}{\pi}\int d\omega\int d\tilde E P(\omega) R^\omega_{E \tilde E},
\label{Iee}
\en
where
\beg
\begin{split}
R^\omega_{E \tilde E}=f_{E} f_{\tilde E-\omega} (1- f_{E-\omega} )(1-f_{\tilde E} )\\
- f_{E-\omega}  f_{\tilde E} (1-f_{E}) (1-f_{\tilde E-\omega}),
\end{split}
\en
$f_E\equiv f(E,t)$ is the electron distribution function, and $P(\omega)$ is the transition probability. Importantly, $P(\omega)$ is even in $\omega$. The extra factor of $\frac{1}{\pi}$ in \eref{Iee} as compared to the expression in~\cite{rammer_sup} is due to our rescaling $t\to\pi t$. Observe that $I_\mathrm{ee}(E)$ is identically zero for the Fermi distribution $f(E)=[ e^{\frac{E}{T}}+1]^{-1}$ for any $T$.

It is simple to verify the energy conservation, i.e., that
\beg
\int_{-\infty}^\infty \!\!\!dE \, E I_\mathrm{ee}(E)=0.
\en
Linearization requires more work, but it is straightforward algebra, which we do not show here. The end  result is of the form
\beg
\delta I_\mathrm{ee}(E)= \int  \!\! dE\, M^\mathrm{ee}_{E \tilde E} \varphi_{\tilde E}
\en
with manifestly real-symmetric integration kernel $M^\mathrm{ee}_{E \tilde E}$.

\section{Full kinetic equation}

As we discuss in more detail in Sec.~\ref{kin_der} below, the electron-phonon kinetic equation,
\beg
 (1-\Sigma') \dot f +f' \dot\Sigma = I_\mathrm{ep},
\label{ke_sup}
\en 
is derived (as usual for the kinetic framework) under the assumption that the electron distribution function $f(E, t)$ varies in time sufficiently slowly.  

On the other hand, we saw that as we approach the instability from the right and for $\lam>\lam_c$ some of the generalized eigenvalues $\gamma$  diverge (Fig.~\ref{FigEigSpec}).   The corresponding eigenmodes change in time very fast, and the gradient expansion is not valid for them. However, this does not  affect the existence of the instability. First, in a stable system all fast modes quickly decay and its long-time dynamics is still governed by \eref{ke_sup}. If the exact dynamics is stable, we expect \eref{ke_sup} to be stable too. Otherwise, the fast modes  decay until the dynamics is described by \eref{ke_sup}   and then grow until \eref{ke_sup} is no longer valid. Thus, the system  oscillates perpetually instead of relaxing  to the thermal equilibrium. Second, we see from the analysis of the kinetic equation in the strong coupling regime [see, for example, \eref{burger1}] that if the initial deviation from the equilibrium is very smooth in energy (varies little with $E$), its initial unstable dynamics is correspondingly very slow and is therefore accurately described by \eref{ke_sup}.

To simulate the full kinetic equation \eqref{ke_sup}, we use the same discrete energy grid as in Sec.~\ref{linear_sup}. This converts \eref{ke_sup} into a differential algebraic equation (DAE) of the form
\begin{equation}
\mathbb{G}\left( \dot{\bm{f}}, \bm{f} \right) = 0,
\label{dae_sup}
\end{equation}
where $\bm{f}$ is an $L$-dimensional vector $\{ f(\epsilon, t), f(2\epsilon, t), \dots, f(L\epsilon, t) \}$ and $\dot{\bm{f}}$ denotes the time derivative of $\bm{f}$. To obtain the above equation, we used the particle-hole symmetry and the following boundary condition: 
\begin{equation}
f(0,t) = \frac{1}{2}, \qquad f(j\epsilon, t) = 0, \quad\textrm{ if } j>L.
\label{bc_sup}
\end{equation}
Instead of rewriting the DAE \eqref{dae_sup} as an ordinary differential equation, we directly solved \eref{dae_sup} using the package {\fontfamily{cmtt}\selectfont Sundials.jl}. This wrapper package in turn imports solvers from the {\fontfamily{cmtt}\selectfont SUNDIALS} integrator package \cite{gardner2022sundials, hindmarsh2005sundials} that can be implemented using the {\fontfamily{cmtt}\selectfont Julia} programming language. These solvers are particularly suitable for the kind of nonlinear and stiff differential equation that we have in \eqref{dae_sup}. 

 We integrate \eref{dae_sup} with the boundary condition \eqref{bc_sup} and the initial condition
\beg
\bm{f}(t = 0) = \{ f_{0}(\epsilon, T_\mathrm{in}), f_{0}(2\epsilon, T_\mathrm{in}), \dots, f_{0}(L\epsilon, T_\mathrm{in}) \},
\label{ic_sup}
\en
where $f_{0}$ is the equilibrium Fermi distribution with a temperature $T_\mathrm{in}$ slightly higher than the phonon temperature $T_0$, using the {\fontfamily{cmtt}\selectfont SUNDIALS} DAE solver   {\fontfamily{cmtt}\selectfont IDA} \cite{IDAdoc}. In this algorithm, the time derivatives at the $n^\mathrm{th}$ time step $t_{n}$ are written using the backward differentiation formula  of order $q$ as
\begin{equation}
\dot{\bm{f}}(t_{n}) = \frac{1}{h_{n}} \sum_{i=0}^{q} \alpha_{n,i} \bm{f}(t_{n-i}),
\label{bdf_sup}
\end{equation}
where $h_{n} = t_{n} - t_{n-1}$, $q$ ranges from 1 to 5, and the coefficients $\alpha_{n,i}$ are uniquely determined by the order $q$. Applying the above formula to \eref{dae_sup}, we reduce the original DAE problem to the problem of solving a nonlinear algebraic system by Newton's method at each time step. Given the initial condition $\bm{f}(t_0)$, one has to use $q = 1$ to obtain $\bm{f}(t_{1})$. Then the step size $h_n$ and the order $q$ are selected after a local error test \cite{IDAdoc}. For instance, it is possible to use $q = 2$ to calculate $\bm{f}(t_{2})$. 

 As has been mentioned in the main text, see Fig.~\ref{FigInstability}, one ends up having rapidly oscillating solutions -- especially close to the zeros of $Z_0^{-1}(E)$ -- when $\lambda > \lambda_c$. This is also seen when one numerically simulates the simplified kinetic equation~\re{burger1} in the strong coupling limit. As a result, after a few time steps, the {\fontfamily{cmtt}\selectfont IDA} integrator fails the local error test too many times \cite{IDAdoc, DAEBook}. This makes either $q>q_\mathrm{max}$ or $h_{n}<h_\mathrm{min}$. At this point, {\fontfamily{cmtt}\selectfont IDA} stops and returns  a give-up message. If $\lambda < \lambda_c$, the simulation of \eref{ke_sup} runs until the provided end-times and obtains
\beg
\bm{f}(t = t_\mathrm{end}) = \{ f_{0}(\epsilon, T_0), f_{0}(2\epsilon, T_0), \dots, f_{0}(L\epsilon, T_0) \}.
\label{asymp_eq_sup}
\en

 \section{Kinetic equation: short review}
\label{kin_der}

The electron-phonon kinetic equation~\re{ke_sup} employed in this work forms the backbone of the theory of nonequilibrium phenomena in metals, playing a role analogous to that of Maxwell's equations in electromagnetism.  
In its modern formulation, it is derived from the Dyson equation in the Keldysh formalism~\cite{keldysh1965,rammer_sup} under two principal assumptions:  
(a) the ions are much slower than the electrons, i.e., the maximum phonon frequency $\Omega \ll E_F$, and  
(b) the electron distribution function $f(E,t)$ varies sufficiently slowly in time.  
These conditions justify a gradient expansion and the use of Migdal--Eliashberg theory, even for strong renormalized electron-phonon coupling $\lambda \gtrsim 1$, as long as $\Omega / E_F \ll 1$~\cite{migdal_sup,baym1961,prange1_sup}.

The first step is a gradient expansion of the Dyson equation in the Wigner representation with respect to relative frequency and momentum. Assuming the so-called parameter of the kinetic equation, $\tau_f E_\Sigma \gg 1$, where $\tau_f$ and $E_\Sigma$ are the characteristic time and energy scales over which $f$ and the self-energy $\Sigma$ vary, it is justified to retain only the leading term --- the Poisson bracket  
\begin{equation}
\{f, \Sigma\} = f' \dot{\Sigma} - \Sigma' \dot{f},
\end{equation}  
on the left-hand side of the kinetic equation~\cite{baym1961}.

Next, one integrates over the single-particle energy $\xi_{\bm p}$. In the limit $\Omega/E_F \to 0$, the self-energy $\Sigma$ becomes momentum-independent, and the electron-phonon interaction scatters electrons from one point on the Fermi surface to another. Then, the summation over the phonon momentum $\bm q$ in the electron-phonon collision integral reduces to  
\begin{equation}
\sum_{\bm q} \frac{g_{\bm q}^2}{2\omega_{\bm q}} \delta(\omega - \omega_{\bm q}) \int \frac{d\Omega'}{4\pi} \delta(\bm k_F - \bm k'_F - \bm q) \equiv \alpha^2 F(\omega),
\end{equation}  
where $g_{\bm q}$ is the electron-phonon matrix element and $\omega_{\bm q}$ is the phonon frequency~\cite{allendynes_sup,mitrovic_sup,carbotte_sup,2008review_sup}. The left-hand side defines the Eliashberg spectral function $\alpha^2 F(\omega)$.

The kinetic equation then takes the form  
\begin{equation}
(1 - \Sigma') \dot{f} + f' \dot{\Sigma} + \bm v_F \cdot \nabla f = I_{\mathrm{ep}},
\label{kin111}
\end{equation}  
where $I_{\mathrm{ep}}$ is the electron-phonon collision integral:
\begin{equation}
\begin{split}
I_{\mathrm{ep}} = 2\pi \! \int_0^\infty \!\! d\omega\, \alpha^2 F(\omega) \left[ N(\omega)(f_+ + f_- - 2f) -\right.\\ 
\left. -f(f_+ - f_-) + f_+ - f \right],
\end{split}
\end{equation}  
with $f_\pm = f(E \pm \omega)$ and $N(\omega)$ the phonon distribution function.  
Note that the dimensionful electron-phonon coupling $g_{\bm q}$ is averaged over the Fermi surface in $\alpha^2 F(\omega)$; thus, its detailed momentum dependence does not affect the dynamics of the electron distribution to leading order in $\Omega / E_F$~\cite{rammer_sup,keldysh1965}.

This kinetic equation has been derived within the conserving approximation framework~\cite{baym1961,rammer_sup}, and ensures energy and particle number conservation when $\Sigma$ is determined self-consistently. In equilibrium, the Fermi--Dirac distribution $f_0(E,T_0)$ is a stationary solution, and the quasiparticle weight $Z_0^{-1}(E) = 1 - \Sigma'[f_0]$ emerges naturally from the formalism.

 It is important to emphasize that our kinetic approach bypasses the issue of phonon renormalization within electron-phonon models (see the introductory part of the main text). The kinetic equation treats the physical phonon spectrum $\omega_{\bm q}$ and the dimensionful electron-phonon coupling $g_{\bm q}$ as inputs. The goal of our work is to establish a stability constraint on these quantities, irrespective of how they are obtained --- whether from experiment, ab initio calculations, or self-consistently from an electron-phonon Hamiltonian, as in Prange and Kadanoff (1964)~\cite{prange1_sup}.

On the timescale relevant to the instability, the phonon subsystem remains close to equilibrium (see below), so dynamical phonon feedback is negligible. Accordingly, there is no need for a separate kinetic equation for phonons: the influence of lattice dynamics is already encoded in the renormalized inputs. This standard kinetic treatment is well-suited to our purpose, which is to determine the onset of instability as a function of the physical coupling strength, regardless of how that coupling is determined.

In this work, we consider spatially uniform deviations of the electron distribution function from equilibrium. Then, the spatial gradient term $\bm v_F \cdot \nabla f$ on the left-hand side vanishes, and the distribution evolves according to Eq.~\re{ke_sup}.  

Recall that in our proof that $C_{\mathrm{el}} < 0$ is a sufficient condition for instability, we assumed the initial electron distribution was the (spatially uniform) thermal distribution at temperature $T_{\mathrm{in}}$ slightly above the lattice temperature $T_0$. In linear stability analysis, the specific initial condition is irrelevant: the system is unstable as long as there exists a single unstable normal mode, and stable otherwise.

Even though the initial dynamics of the instability are spatially uniform (as in, e.g., the superconducting instability, which is also translationally invariant), spatial structure can emerge at later times due to coupling between degrees of freedom at different length scales, a common feature in nonlinear media. 

In charge density wave systems, for example, a Fermi-surface instability leads to a periodic modulation of both the electronic charge and the lattice distortion~\cite{gruner1988}. In high-pressure hydrides, such as H$_3$S, LaH$_{10}$, and YH$_9$, strong electron-phonon coupling can induce dynamical instabilities in high-symmetry phases that resolve into distorted or modulated structures~\cite{drozdov1_sup,errea2016,minkov1_sup,chen_sup,kong_sup,bi1_sup,chen3_sup,semenok_sup}. Similarly, post-quench dynamics in superconductors can exhibit the spontaneous formation of spatial inhomogeneities due to parametric excitation of pairing modes and the onset of Cooper pair turbulence~\cite{dzero,chern,bulgac2020,yang,fan}. These effects are often understood as secondary instabilities or modulational instabilities, and are sometimes described by wave turbulence phenomenology~\cite{zakharov1992}. 

The investigation of such late-time spatial structure lies beyond the scope of this work. Our goal here is to demonstrate the existence of the instability itself. Since its onset is spatially uniform, it is sufficient for our purposes to analyze the dynamics in the homogeneous setting. The nonlinear spatial evolution following the instability is an interesting direction for future research.

Finally, we comment on the widely used equilibrium phonon approximation, which we have also adopted in this work. This approximation is well justified by the separation of energy and time scales between the electron and phonon subsystems~\cite{phys_kin_sup,abrikosov_sup,allen_1987,eli_dynamics,rammer_sup}. At relevant temperatures, the phonon specific heat exceeds the electronic specific heat by a large factor, $C_{\mathrm{ph}}/|C_{\mathrm{el}}| \sim E_F/\Omega$, which typically ranges from $10^2$ to $10^4$ in metals. As a result, the lattice can absorb energy from electrons with minimal change in its distribution and effectively serves as a thermal reservoir. In addition, the phonon-electron collision integral --- responsible for driving phonons out of equilibrium --- is smaller than the electron-phonon one by a factor of $T/E_F \sim \Omega/E_F$~\cite{phys_kin_sup}, making the dynamics of phonon distribution  much slower. Consequently, it remains near the initial equilibrium on the timescale relevant to the instability.

Effects that fall outside the equilibrium phonon approximation, such as phonon drag (which is parametrically small in metals~\cite{section81}) or phonon bottlenecks in polar semiconductors~\cite{zou2023}, are irrelevant in our context. Moreover, although the phonon distribution may evolve at later times --- for instance, through heating or structural transitions in hydrides --- such processes occur only after the instability has already developed. Since our goal is to establish the existence of the instability and analyze its early-time dynamics, the equilibrium phonon approximation remains fully adequate for our purposes.

\vspace{10pt}

 \section{Benchmarking the kinetic equation}

Because our main results are based on the electron-phonon kinetic equation~\re{ke_sup} it is important to  put our study into the context of other  applications of this equation.
In particular, we must demonstrate that the kinetic equation accurately captures phenomena where the quasiparticle weight \( Z^{-1} = 1-\Sigma' \) plays a central role, as it does in our instability analysis. To this end, we review the derivation of the kinetic equation~\re{ke_sup} and key benchmark cases where it has been successfully tested against experiment and alternative theoretical approaches.

Applications of the kinetic equation where the quasiparticle weight \( Z^{-1} \) plays a central role are numerous and span a wide range of materials, including normal metals, semiconductors, and superconductors. Both the structure of the kinetic equation and the critical role of the quasiparticle weight have been extensively validated through comparisons with experimental data and alternative theoretical approaches. Below, we highlight several representative examples where these validations are particularly direct and relevant to the present work.

One of the earliest explicit uses of the kinetic equation was by Prange and Kadanoff in 1964 in their study of normal metals~\cite{prange1_sup}. Using this equation, they demonstrated that steady-state transport coefficients (such as electrical and thermal conductivities and spin diffusion) are unaffected by electron-phonon many-body corrections to leading order. This explained why conventional Boltzmann theory\textemdash which neglects the energy-time Poisson bracket, \( \{f, \Sigma\} = f' \dot{\Sigma} - \Sigma' \dot{f} \)\textemdash successfully accounted for experimental transport measurements, thereby validating its consistency with many-body theory. This provides an early and important confirmation that the kinetic equation correctly incorporates key many-body effects, including the quasiparticle weight.

In the same work, Prange and Kadanoff also derived the expression for the electronic specific heat, which matches experimental results within a few percent accuracy~\cite{2008review,carbotte,weird,golubov}. Notably, this expression directly follows from the left-hand side of the kinetic equation, and the quasiparticle weight, together with the \( \Sigma' \dot{f} \) term, plays an essential role. Without these contributions, the specific heat would simply correspond to that of free electrons, inconsistent with experimental observations. Thus, the kinetic equation captures the essential renormalization effects responsible for the accurate description of thermodynamic properties.

Another benchmark is the high-field Nernst-Ettingshausen effect, which relates the current density to the vector product of the temperature gradient and the magnetic field. The kinetic equation (with a minor modification to include magnetic field terms) predicts the enhancement of the Nernst-Ettingshausen coefficient by a factor of \( Z^{-1} = 1-\Sigma' \), in good agreement with thermopower measurements.

A third important example is the derivation by Allen~\cite{allen_1987} of the hot-electron relaxation rate in metals from the kinetic equation, explicitly retaining the quasiparticle renormalization via the electron-phonon coupling constant \( \lambda = -\Sigma'|_{E=0} \). His prediction that the relaxation time is proportional to \( \lambda \langle \omega^2 \rangle \), where \( \langle \omega^2 \rangle \) is the average squared phonon frequency, was subsequently confirmed by ultrafast pump-probe experiments~\cite{brorson_1990}. These experiments provided a new and accurate method to measure the electron-phonon coupling strength in metals, once again validating the predictive power of the kinetic equation.

Taken together, these examples demonstrate that the kinetic equation reliably captures the key physical phenomena associated with electron-phonon interactions, even when quasiparticle weight renormalization plays a decisive role. Effects originating from the quasiparticle weight, such as those described above and the instability uncovered in our work, are therefore not artifacts of the kinetic theory.

Therefore, the instability we uncover reflects a genuine physical effect, not a shortcoming of the kinetic equation. The logic of our proof is as follows. Assuming the system is a stable metal with a large Fermi energy, the kinetic equation must apply as the appropriate low-energy description. However, it then follows from the kinetic equation that the system becomes unstable once the renormalized electron-phonon coupling exceeds a certain threshold. This establishes that a metallic state with such a coupling strength cannot be realized physically. The conclusion follows by proof through contradiction.

\section{Strong coupling regime}

Let us estimate the energy uncertainty $\Delta E$ in the strong coupling regime $T\gg\Omega$ for Einstein phonons with frequency $\Omega$. This uncertainty is of the order of the inverse dephasing time $\tau_\phi$. The electron-phonon collision time in this regime is $\tau\sim (\lam T)^{-1}$, see, e.g.,~\cite{mitrovic_sup}. The change in electron energy in each collision is very small, of the order of $\Omega$.  Modeling the electron energy dynamics as a 1D random walk with a step $\Omega$, we estimate  that over the dephasing time the energy changes by
\beg
\Delta E\sim \Omega \sqrt{\frac{\tau_\phi}{\tau} }.
\en
Since, on the other hand, $\Delta E\sim \tau_\phi^{-1}$, equating these two expressions, we obtain
\beg
\Delta E\sim (g^2 T)^{1/3}.
\label{uncertainty}
\en
Since the Eliashberg theory is independent of the phonon spectrum in the strong coupling limit, we expect this formula for $\Delta E$ to hold for general phonon dispersion, not just Einstein phonons.

The expression for $Z_0^{-1}$ in the strong coupling limit obtains from \eref{zfull} by taking the limit $\omega\to 0$ under the integral. We have
\beg
\frac{1}{Z_0(E)}=1 -\frac{g^2}{4\pi^2 T_0^2} \Re\left[ \psi^{(2)}\left(\frac{1}{2}+\frac{iE}{2\pi T_0}\right)\right].
\en
Here and in \eref{uncertainty} we restored $g$ for clarity (recall that we are  using the energy units where $g=1$).
The length of the interval where $Z_0^{-1}<0$ is $\delta E= (1.16-0.40)g=0.76 g$, which is parametrically larger than the energy uncertainty~\re{uncertainty} at temperatures $T\ll g$. Already at $T=T_c=0.1827 g$, we have $\delta E> (g^2 T)^{1/3}=0.57 g$.


\begin{thebibliography}{99}
\bibitem{polaron} In this context,  a polaronic or bipolaronic transition is a phase change involving charge localization and lattice distortion~\cite{ZSRBMPS,Chakraverty,alexandrov,capone,Nosarzewski}; at low temperatures, this often leads to charge ordering analogous to a CDW.
\end{thebibliography}
\end{document}